\documentclass[pra,aps,amsmath,amssymb,amsfonts,twocolumn,nofootinbib]{revtex4}
\usepackage{amssymb}
\usepackage{}
\usepackage{bm,mathrsfs}

\usepackage{graphicx}
\usepackage{epsfig}
\usepackage{amsmath,bbm}
\usepackage{amsfonts,amssymb}
\usepackage{times}
\usepackage{verbatim}
\usepackage[sort&compress]{natbib}

\usepackage{amsmath}
\usepackage[colorlinks,breaklinks,linkcolor=blue,anchorcolor=blue,citecolor=blue,urlcolor=blue,
dvipdfm
]{hyperref}

\newcommand{\be}{\begin{equation}}
\newcommand{\ee}{\end{equation}}
\newcommand{\bea}{\begin{eqnarray}}
\newcommand{\eea}{\end{eqnarray}}

\def\>{\rangle}
\def\<{\langle}

\def\qed{\leavevmode\unskip\penalty9999 \hbox{}\nobreak\hfill
     \quad\hbox{\leavevmode  \hbox to.77778em{%
               \hfil\vrule   \vbox to.675em%
               {\hrule width.6em\vfil\hrule}\vrule\hfil}}
     \par\vskip3pt}
     
\begin{document}

\newtheorem{theorem}{Theorem}
\newtheorem{lemma}[theorem]{Lemma}
\newtheorem{corollary}[theorem]{Corollary}
\newtheorem{proposition}[theorem]{Proposition}
\newtheorem{definition}[theorem]{Definition}
\newtheorem{example}[theorem]{Example}
\newtheorem{conjecture}[theorem]{Conjecture}

\title{Engineering steady-state entanglement via dissipation and quantum Zeno dynamics in optical cavity}

\author{D. X. Li}
\affiliation{Center for Quantum Sciences and School of Physics, Northeast Normal University, Changchun, 130024, People's Republic of China}
\affiliation{Center for Advanced Optoelectronic Functional Materials Research, and Key Laboratory for UV Light-Emitting Materials and Technology
of Ministry of Education, Northeast Normal University, Changchun 130024, China}

\author{X. Q. Shao\footnote{Corresponding author: shaoxq644@nenu.edu.cn}}
\affiliation{Center for Quantum Sciences and School of Physics, Northeast Normal University, Changchun, 130024, People's Republic of China}
\affiliation{Center for Advanced Optoelectronic Functional Materials Research, and Key Laboratory for UV Light-Emitting Materials and Technology
of Ministry of Education, Northeast Normal University, Changchun 130024, China}

\author{J. H. Wu}
\affiliation{Center for Quantum Sciences and School of Physics, Northeast Normal University, Changchun, 130024, People's Republic of China}
\affiliation{Center for Advanced Optoelectronic Functional Materials Research, and Key Laboratory for UV Light-Emitting Materials and Technology
of Ministry of Education, Northeast Normal University, Changchun 130024, China}

\author{X. X. Yi}
\affiliation{Center for Quantum Sciences and School of Physics, Northeast Normal University, Changchun, 130024, People's Republic of China}
\affiliation{Center for Advanced Optoelectronic Functional Materials Research, and Key Laboratory for UV Light-Emitting Materials and Technology
of Ministry of Education, Northeast Normal University, Changchun 130024, China}
\date{\today}

\begin{abstract}
A new mechanism is proposed for dissipatively preparing maximal Bell entangled state of two atoms in an optical cavity. This scheme integrates the spontaneous emission, the light shift of atoms in the presence of dispersive microwave field, and the quantum Zeno dynamics induced by continuous coupling, to obtain a unique steady state irrespective of initial state.  Even for a large cavity decay, a high-fidelity entangled state is achievable at a short convergence time, since the occupation of cavity mode is inhibited by the Zeno requirement. Therefore, a low single-atom cooperativity $C=g^2/(\kappa\gamma)$ is good enough for realizing a high fidelity of entanglement in a wide range of decoherence parameters.  As a
straightforward extension, the feasibility for preparation of two-atom Knill-Laflamme-Milburn state with the same mechanism is also discussed.
\end{abstract}

\maketitle

 Quantum entanglement, as a fundamental feature in quantum mechanics, plays an important role in the development of quantum information. It describes a strongly correlated system constituted by pairs or groups of particles, where a measurement, made on either of the particles apparently collapses the state of the system instantaneously \cite{pra022317ref1}. Because of the effectiveness and security, the quantum entangled states have been used in various practical applications, ranging from quantum teleportation to one-way quantum computation \cite{bouwmeester1997experimental,pra064302ref3,PhysRevLett.86.5188,walther2005experimental}.  And then, more and more interests are devoted to study the preparation of quantum entanglement, especially in the field of experimental quantum information science \cite{pra022317ref9,pra022317ref5}.

As is well known, the quantum dissipation induced by the coupling between quantum systems and the environment is a major issue in the development of quantum science and technology, which is mathematically characterized by a Lindblad generator in Markovian quantum master equations. Traditionally, it has been considered that the dissipation can only have detrimental effects in quantum information processing tasks because of the decoherent effect on investigated quantum system. However, recent studies have changed the view of people for dissipation due to the fact that the environment can be used as a resource for quantum computation and entanglement generation \cite{pra012319ref13,pra012319ref14,pra012319ref20,pra012319ref19,PhysRevA.89.012319,Shen:14,PhysRevA.95.062339}.
In particular, G. Vacanti and A. Beige \cite{pra012319ref13} discussed the possibility of preparing highly entangled states via spontaneous emission from excited atomic states. In Ref.~\cite{PhysRevA.89.012319}, Shao \textit{et al.} proposed a dissipative scheme to prepare a three-dimensional entanglement in an optical cavity, where the atomic decay played a positive role in the dynamic evolution.

The quantum Zeno effect is an efficient way of combating decoherence. Under frequent measurement, an unstable particle can never decay \cite{jpbref21}. The first experimental observation of the quantum Zeno effect  was achieved by Cook \cite{jpbref22}. And then the effect was successfully demonstrated in experiments for different systems, such as photon polarization \cite{jpbref23}, cavity
QED \cite{jpbref24} and Bose-Einstein condensates \cite{jpbref25}, and so on.  However, it is not necessary for the quantum Zeno effect to freeze everything and more general phenomena can take place. The quantum Zeno dynamics \cite{Beige2000b,Facchi2000,jpbref26} allows the system to evolve away from its initial state under a multidimensional projection. In the context of cavity quantum electrodynamics, the quantum Zeno dynamics induced by a strongly continuous coupling guarantees a robust way of quantum information processing against cavity decay \cite{pra213601,pra032120,nphys3076,Bretheau776,Barontini1317,PhysRevLett117140502}.

Rigorous extensions of the quantum Zeno dynamics have appeared in Ref.~\cite{Beige2000b,Facchi2000,jpbref26}. Here, we only give an elementary introduction to the quantum Zeno dynamics induced by continuous coupling. A generic Hamiltonian of a dynamical evolution can be written as $H_K=H+KH_c,$ where $H$ is the Hamiltonian of the quantum system, $H_c$ is an additional interaction Hamiltonian caricaturing the continuous measurement and $K$ is coupling constant. In the limit of $K\rightarrow\infty$, the evolution operator of the system, $U_K(t)=\exp{(-iH_Kt)}$, is dominated by $\exp{(-iKH_ct)}$. Then we can consider the limiting operator as $U_Z(t)=\lim_{K\rightarrow\infty}\exp{(iKH_ct)}U_K(t),$ which can be rewritten as $U_Z(t)=\exp{(-iH_Zt)},$
where
$H_Z=\hat PH=\sum_nP_nHP_n$
is the Zeno Hamiltonian, and $P_n$ is the eigenprojection of $H_c$ with corresponding eigenvalue $\eta_n$, in other words,
$H_c=\sum_n\eta_nP_n.$
In conclusion, the limiting evolution operator of the whole system can be written as,
$U_K(t)=\exp{\Big( -i\sum_nKt\eta_nP_n+P_nHP_nt \Big)},$
which is the basic formula to obtain the effective Hamiltonian in the next section.

\begin{figure}
\begin{minipage}[t]{0.49\linewidth}
\centering
\includegraphics[scale=0.08]{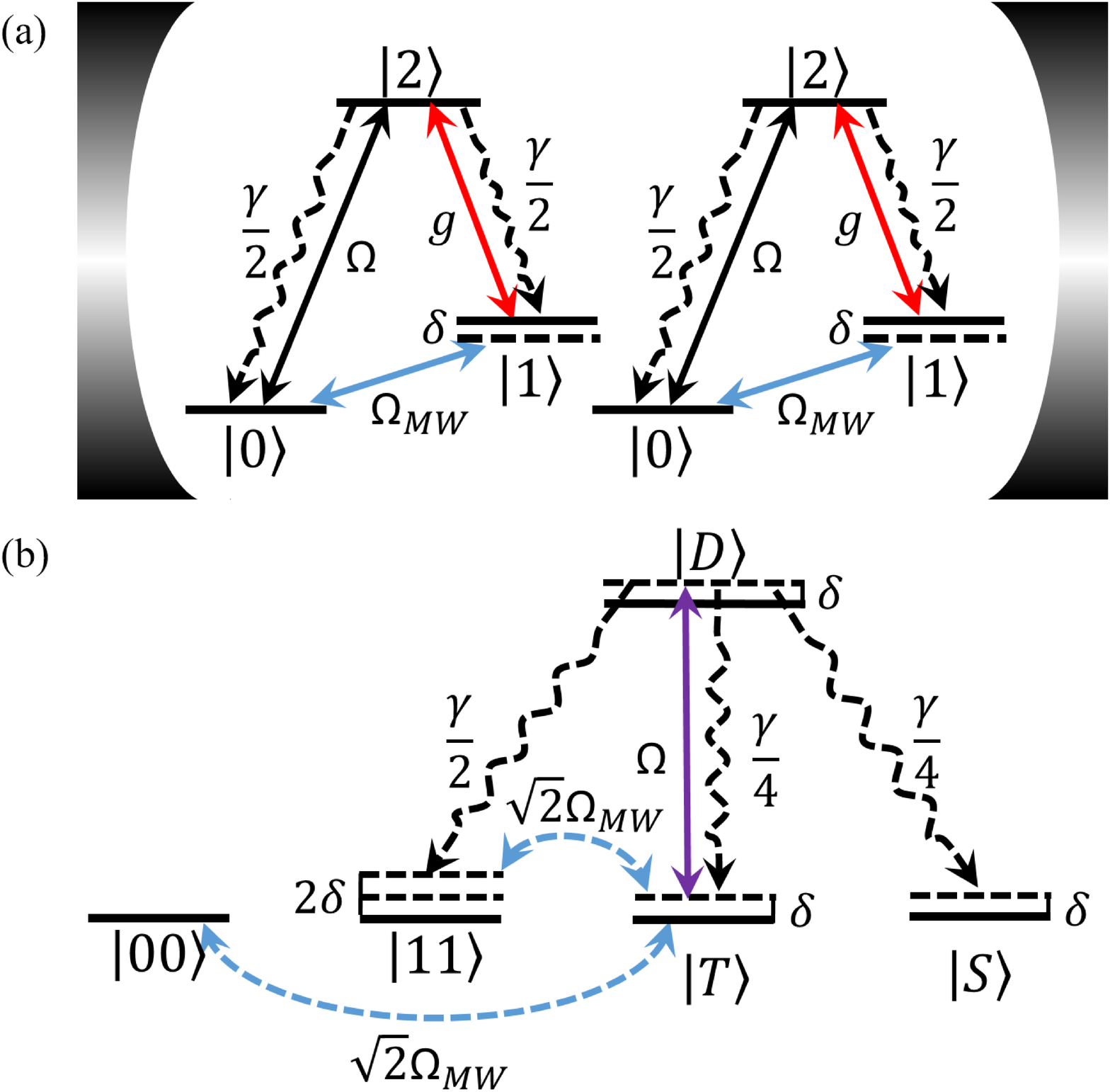}
 \centerline{}
\end{minipage}
\begin{minipage}[t]{0.49\linewidth}
\centering
\includegraphics[scale=0.30]{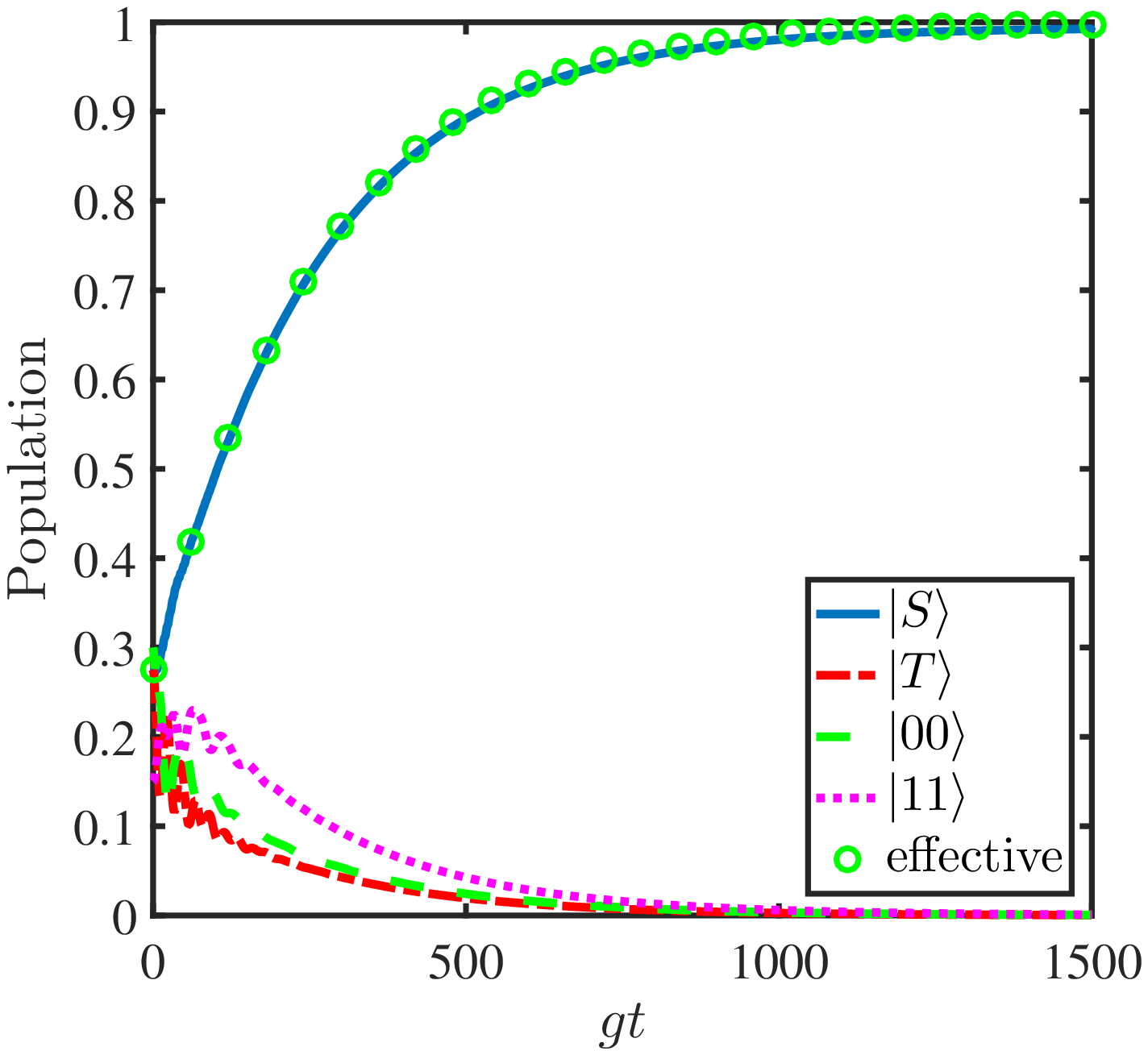}
 \centerline{(c)}
\end{minipage}
\caption{(Color online) (a) Level diagram of two $\Lambda$-type atoms A and B trapped into an optical cavity. The $\gamma$ and $\kappa$ denote the rate of atomic decay and cavity decay, respectively.  (b) Effective transitions of two atoms. (c) Populations of corresponding states are plotted as functions of $gt$.  }
\label{model}
\end{figure}

Our setup consisting of two $\Lambda$ atoms trapped in an optical cavity is shown in Fig.~\ref{model}(a). The transition between states $|1\rangle$ and $|2\rangle$ is coupled to the quantized cavity field resonantly with coupling strength $g$, and the transition between states $|0\rangle$ and $|2\rangle$ is driven a resonant laser with Rabi frequency $\Omega$. Besides, the transition between ground states $|0\rangle$ and $|1\rangle$ is caused by a microwave field with Rabi frequency $\Omega_{MW}$, detuning $\delta$. Then, we can write down the Hamiltonian of system in the Schr\"{o}dinger picture ($\hbar=1$).
\begin{eqnarray}\label{HS}
H_S=H_0^S+H_I^S+H_{MW}^S,
\end{eqnarray}
where $H_0^S=\sum_{i=A}^B\sum_{j=0}^{2}\omega_j|j\rangle_i\langle j|+\omega_aa^\dag a$, $H_I^S=\sum_{i=A}^Bg|2\rangle_i\langle1|a+\Omega(|2\rangle_A\langle0|+e^{i\phi}|2\rangle_B\langle0|)e^{-i\omega_bt}+{\rm H.c.}$, $H_{MW}^S=\Omega_{MW}\sum_{i=A}^B|1\rangle_i\langle0|e^{-i\omega t}+{\rm H.c.},$
and $a$ and $a^\dag$ denote the annihilation and creation operators of the quantized cavity mode. $\omega_j (j=0,1,2)$ and $\omega_a$ are the corresponding frequencies of state $|j\rangle$ and cavity mode, respectively. $\omega_{b}$ denotes the frequency of the resonant classical field and $\omega$ is the frequency of the dispersive microwave field. $\phi$ stands for the phase difference of the classical optical lasers acting on two atoms and we consider $\phi=\pi$ in the following. Now we reformulate the Hamiltonian in the interaction picture as
$H_I=H_M^I+H_W^I$. And $H_M^I=\sum_{i=A}^B\Omega_{MW}(|1\rangle_i\langle0|+{\rm H.c.})+\delta|1\rangle_i\langle1|-\delta a^\dag a, H_W^I=\sum_{i=A}^Bg|2\rangle_i\langle1|a+\Omega(|2\rangle_A\langle0|-|2\rangle_B\langle0|)+{\rm H.c.},$
and $\delta=\omega_1-\omega_0-\omega$ is the detuning parameter of microwave field. The corresponding dynamics of current system is governed by the master equation
\begin{equation}\label{master}
\dot \rho=-i[H_I,\rho]+\sum_{k=1}^5L^k\rho L^{k\dag}-\frac{1}{2}(L^{k\dag}L^k\rho+\rho L^{k\dag}L^k),
\end{equation}
where the Lindbald operators associated with atomic spontaneous emission and cavity decay are $L^{1(2)}=\sqrt{\gamma/2}|0(1)\rangle_A\langle2|$, $L^{3(4)}=\sqrt{\gamma/2}|0(1)\rangle_B\langle2|$, and $L^5=\sqrt{\kappa}a$, provided that the branching ratio of the atomic decay from state $|2\rangle$ to state $|0(1)\rangle$ is $\gamma/2$ for simplicity.

Referring to the formula of quantum Zeno dynamics, the Hamiltonian $H_W^I$ is able to be written as $H_W^I=\Omega(H_c+KH_p)$, where $K=g/\Omega$, $H_c$ represents the interaction between atoms and classical field, and $H_p$ stands for the interaction between atoms and the cavity field. In the regime of strong coupling $g\gg \Omega$, the limiting condition of the quantum Zeno dynamics, $K\rightarrow\infty$, is satisfied and we can simplify the term $H_W^I$ as
$H_W^I=\Omega|T\rangle\langle D|\otimes|0\rangle_c\langle0|+{\rm H.c.},$
where $|0\rangle_c$ denotes the vacuum state of cavity mode and
$|D\rangle=(|21\rangle-|12\rangle)/\sqrt{2},|T\rangle=(|01\rangle+|10\rangle)/\sqrt{2}.$
Then the total Hamiltonian $H_I$ can be expressed as
\begin{eqnarray}\label{Heff}
H_{\rm{eff}}&=&\sqrt{2}\Omega_{MW}(|00\rangle\langle T|+|11\rangle\langle T|)+\Omega|D\rangle\langle T|+\rm{H.c.}\nonumber\\
&&+\delta(|D\rangle\langle D|+|T\rangle\langle T|+|S\rangle\langle S|+2|11\rangle\langle11|).
\end{eqnarray}
From Eq.~(\ref{Heff}), we can find that the cavity field has been decoupled to the effective model. Thus the cavity decay has no effect on our proposal, and the Lindblad operators describing the spontaneous emission of atoms are in form of
$
L^{1}=\sqrt{\gamma/8}(|T\rangle+|S\rangle)\langle D|,
L^{2}=\sqrt{\gamma/4}|11\rangle\langle D|,
L^{3}=\sqrt{\gamma/8}(|T\rangle-|S\rangle)\langle D|,
L^{4}=\sqrt{\gamma/4}|11\rangle\langle D|,
$
where $|S\rangle=(|01\rangle-|10\rangle)/{\sqrt{2}}$ is the target state to be prepared. Consequently,  the Lindblad terms of Eq.~(\ref{master}) can be rewritten as
\begin{eqnarray}\label{effmasL}
\mathcal{L}\rho&=&\frac{\gamma}{2}|11\rangle\langle D|\rho|D\rangle\langle 11|-\frac{\gamma}{2}(|D\rangle\langle D|\rho+\rho|D\rangle\langle D|)\nonumber\\
&&+\frac{\gamma}{4}|T\rangle\langle D|\rho|D\rangle\langle T|+\frac{\gamma}{4}|S\rangle\langle D|\rho|D\rangle\langle S|.
\end{eqnarray}
From Eqs.~(\ref{Heff}) and (\ref{effmasL}), we can achieve the effective master equation of system
\begin{equation}\label{effmaster}
\dot \rho=-i[H_{\rm eff},\rho]+\sum_{m=1}^3L_{{\rm eff}}^m\rho L_{{\rm eff}}^{m\dag}-\frac{1}{2}(L_{{\rm eff}}^{m\dag}L_{{\rm eff}}^m\rho+\rho L_{{\rm eff}}^{m\dag}L_{{\rm eff}}^m),
\end{equation}
where
$L_{{\rm eff}}^1=\sqrt{\gamma/2}|11\rangle\langle D|,\
L_{{\rm eff}}^{2(3)}=\sqrt{\gamma/4}|T(S)\rangle\langle D|.$

From the effective master equation of Eq.~(\ref{effmaster}), it is intuitive to analytically investigate the evolution of the reduced system, as shown in Fig.~\ref{model}(b). There are four ground states $|00\rangle$, $|T\rangle$, $|11\rangle$ and $|S\rangle$, where the transitions among states $|00\rangle$, $|T\rangle$, and $|11\rangle$ are coupled by the microwave field. Since the optical pumping laser of Rabi frequency $\Omega$ drives the transition $|T\rangle\leftrightarrow|D\rangle$, a quantum state initialized in subspace $\{|00\rangle, |T\rangle, |11\rangle\}$ will always be pumped into the excited state $|D\rangle$ and further spontaneously decay into states $|11\rangle$, $|T\rangle$ and $|S\rangle$, respectively. In the absence of the detuning of microwave field, there is no steady solution of Eq.~(\ref{effmaster}) because the states $(|00\rangle-|11\rangle)/\sqrt2$ and $|S\rangle$ are both the steady states of the system. Once the the detuning of microwave field is introduced, the degeneracy is eliminated and the state $|S\rangle$ becomes the unique steady state of system. Therefore, the cooperation of dissipation, quantum Zeno dynamics, and the detuning of microwave field finally stabilizes the system into the target state $|S\rangle$.

In order to confirm the above analysis, we utilize the definition of population $P=\langle S|\rho(t)|S\rangle$ to assess the performance of current scheme in Fig.~\ref{model}(c). The initial state is randomly selected as $\rho_0=a|00\rangle\langle00|+b|11\rangle\langle11|+c|10\rangle\langle10|+d|01\rangle\langle01|$, with $a=0.3,~b=0.15,~c=0.45$, and $d=0.1$. Under the given parameters $\Omega=0.1g,~\Omega_{MW}=0.5\Omega,~\delta=0.02g,~\gamma=0.1g,~\kappa=0$,  the populations of states $|T\rangle$ (dash-dotted line), $|00\rangle$ (dashed line), and $|11\rangle$ (dotted line) numerically simulated by the full master equation undergo rapid coherent oscillations with
an envelope decaying, while the population of the target state governed by the effective master equation (empty circle) is in good agreement with the population obtained from the full master equation (solid line), which both approach to the maximal value at $gt=1500$. In other words, the effective master equation provides a good approximation for us to explain the behavior of the full master equation.
\begin{figure}
\centering
\includegraphics[scale=0.33]{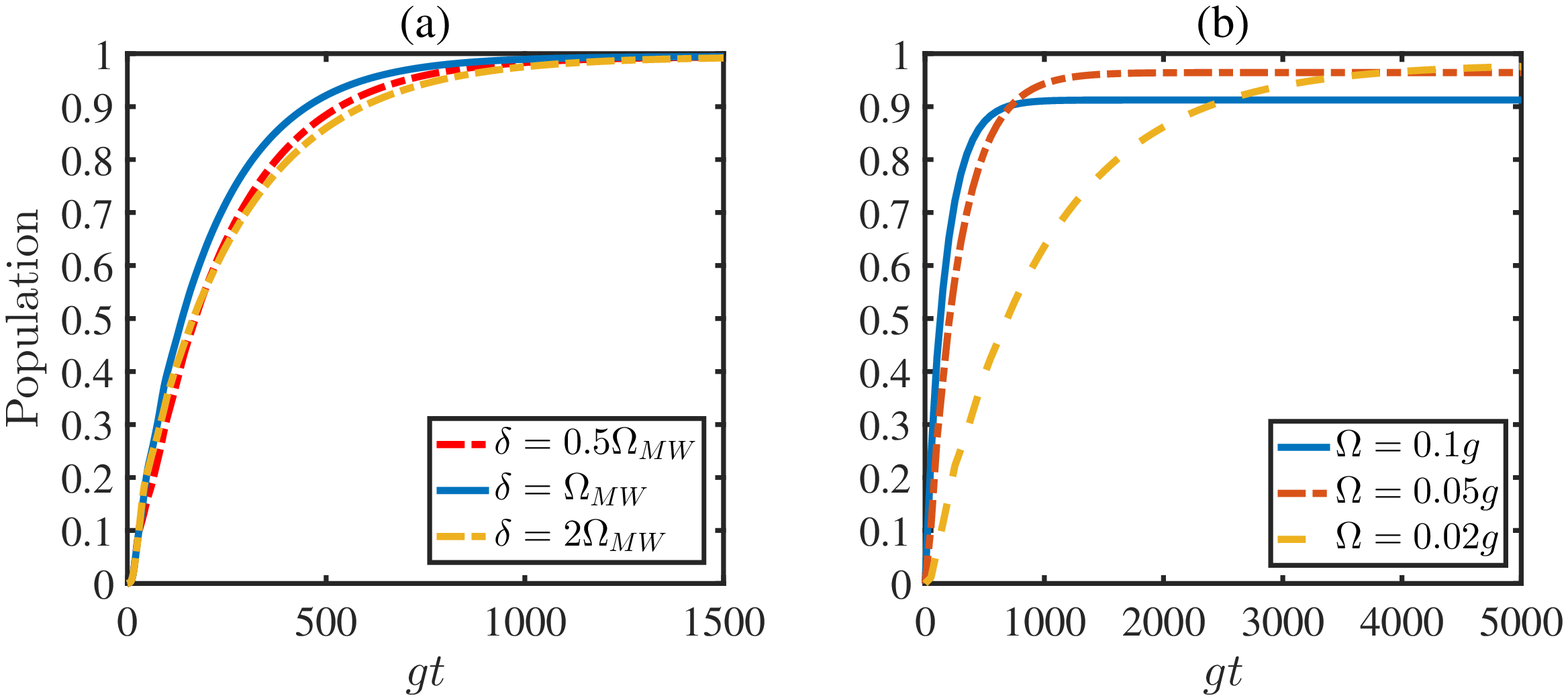}
\caption{(Color online) (a) Populations of $|S\rangle$ with different $\delta$, are plotted as a function of time $gt$. The parameters: $\Omega=0.1g,~\Omega_{MW}=0.5\Omega,~\gamma=0.1g$, and $\kappa=0.$ (b) Dependence of the population for $|S\rangle$ on time is shown for different $\Omega$. Other parameters: $\Omega_{MW}=0.5\Omega,~\gamma=0.1g,~\kappa=0.1g$ and $\delta=\Omega_{MW}$. The initial states both are $|00\rangle$. }
\label{Dchange}
\end{figure}
 In Fig.~\ref{Dchange}(a), we also investigate the influence of the detuning of microwave field on the preparation of entanglement. Starting from the initial state $|00\rangle$, the populations are plotted  by fixing the Rabi frequency $\Omega_{MW}$ but altering the detuning parameter from $\delta=0.5\Omega_{MW}$ to $\delta=2\Omega_{MW}$, while other parameters are same as in Fig.~\ref{model}(c). The result signifies that the current scheme does not rely on a specific value of detuning from the point of view of the steady state \cite{pra012319ref14}. Although there may be an optimal value of $\delta$ leading to a fastest convergence of steady state, the final quality of entanglement is not affected.
\begin{figure}
\centering
\includegraphics[scale=0.35]{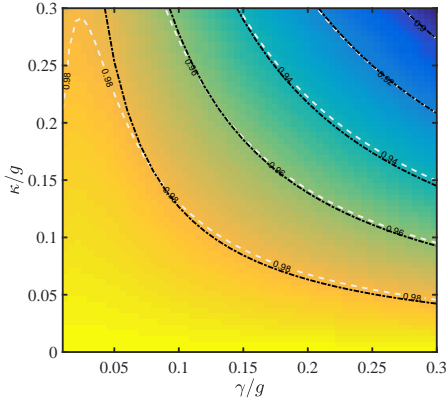}
\caption{(Color online) Contour plot (dashed lines) of the population for $|S\rangle$ in the steady state, and the dashed-dotted curves from left to right correspond to the single-atom
cooperativity parameters $C=79,~36,~23,~16,~\text{and}~12.2$, respectively.  Other parameters:  $\Omega=0.01g$, $\Omega_{MW}=0.5\Omega$ and $\delta=1.3\Omega_{MW}$. }
\label{zong(b)}
\end{figure}

The previous  discussions are mainly based on a model of a perfect cavity. In order to fully characterize the robustness against parameter fluctuation, we consider the cavity decaying at a rate $\kappa$.
 In Fig.~\ref{Dchange}(b), we plot the dependence of the populations on time with different $\Omega$.  According to the limiting condition of quantum Zeno dynamics $K\rightarrow\infty$, the coupling strength $g$ needs to be much larger than the Rabi frequency $\Omega$ in order to have a good approximation for the effective transitions of Fig.~\ref{model}(b). An increase of $\Omega/g$ will destruct the limiting condition, thereby enlarging the deviation between the effective transitions and the realistic transitions, and making the scheme more sensitive to the cavity loss. Then the popolation of desired state will be reduced, accompanied by a shortened convergence time of entanglement.

In Fig.~\ref{zong(b)}, we obtain the steady state $\rho_{s}$ by solving $\dot{\rho}=0$ in Eq.~(\ref{master}), and plot the contour of population for state $|S\rangle$ as functions of atomic spontaneous emission and cavity decay, respectively. It is shown that for a small atomic decay but a big cavity decay, the population is reduced. This behavior exactly proves the practicability of preparing the entangled state $|S\rangle$ by spontaneous emission. However, the population can be still above $90\%$ as long as $0.01g<\gamma<0.27g$, even at a large cavity decay $\kappa=0.3g$ $~(12<C<333)$. This performance verifies again the effective Hamiltonian of quantum Zeno dynamics and the effective transition in Fig.~\ref{model}(b). It also means that a high-fidelity steady entangled state can be prepared for a wide range of decoherence parameters. The dashed-dotted lines denote the cooperativity $C=79,~36,~23,~16,~\text{and}~12.2$ from left to right. Combining the contour plot of population,  we can pick out the optimal values of each cooperativity, which approach $98.15\%$ $(C=79)$, $96.10\%$ $(C=36)$, $94.17\%$ $(C=23)$,  $92.01\%$ $(C=16)$, and $90.00\%$ $(C=12.2)$. This data is instrumental for researchers to seek suitable physical systems for preparation of entanglement with a high fidelity.

\begin{figure}
 \begin{minipage}[t]{0.49\linewidth}
\centering
\includegraphics[scale=0.08]{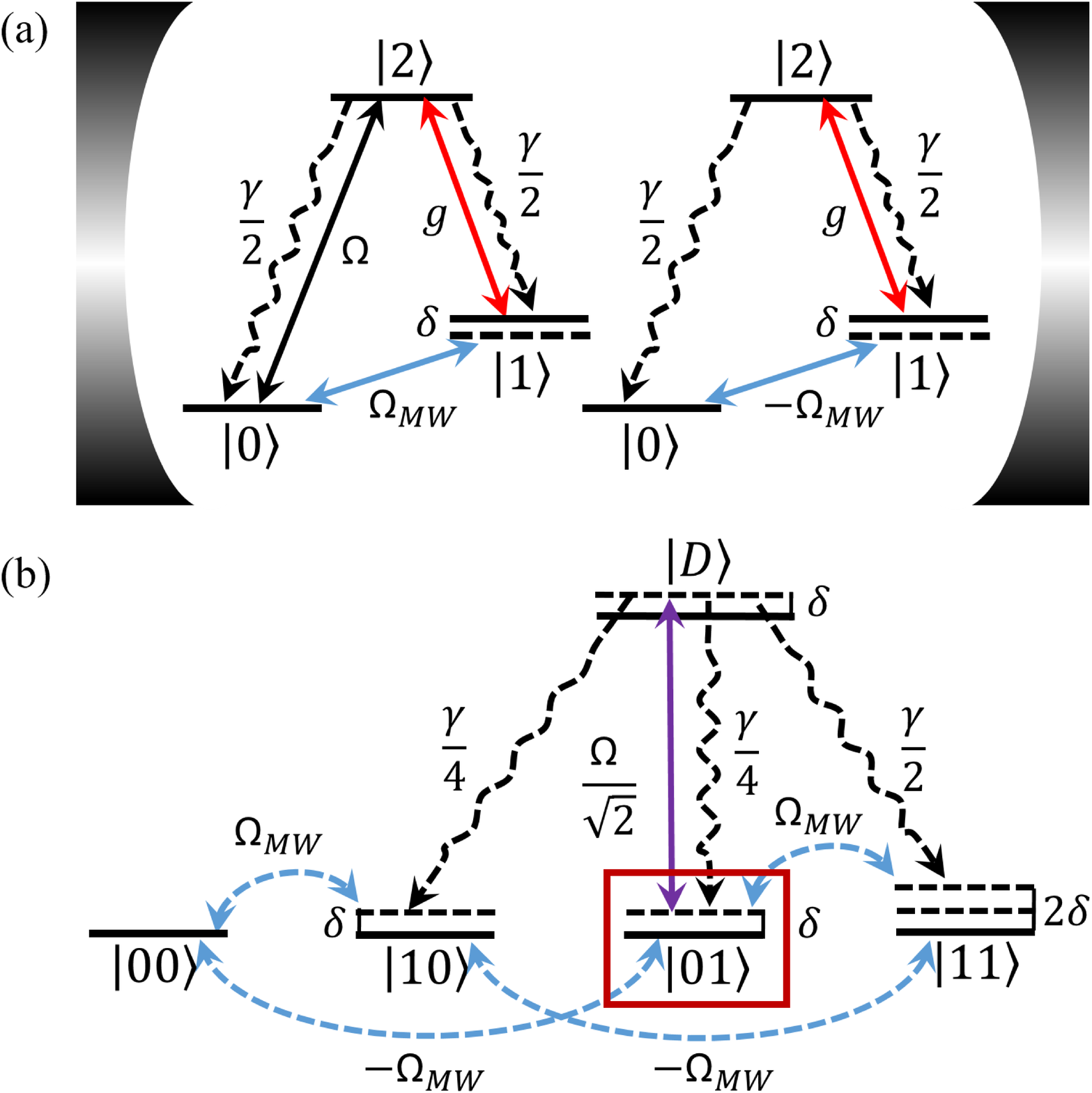}
 \centerline{}
\end{minipage}
\begin{minipage}[t]{0.49\linewidth}
\centering
\includegraphics[scale=0.30]{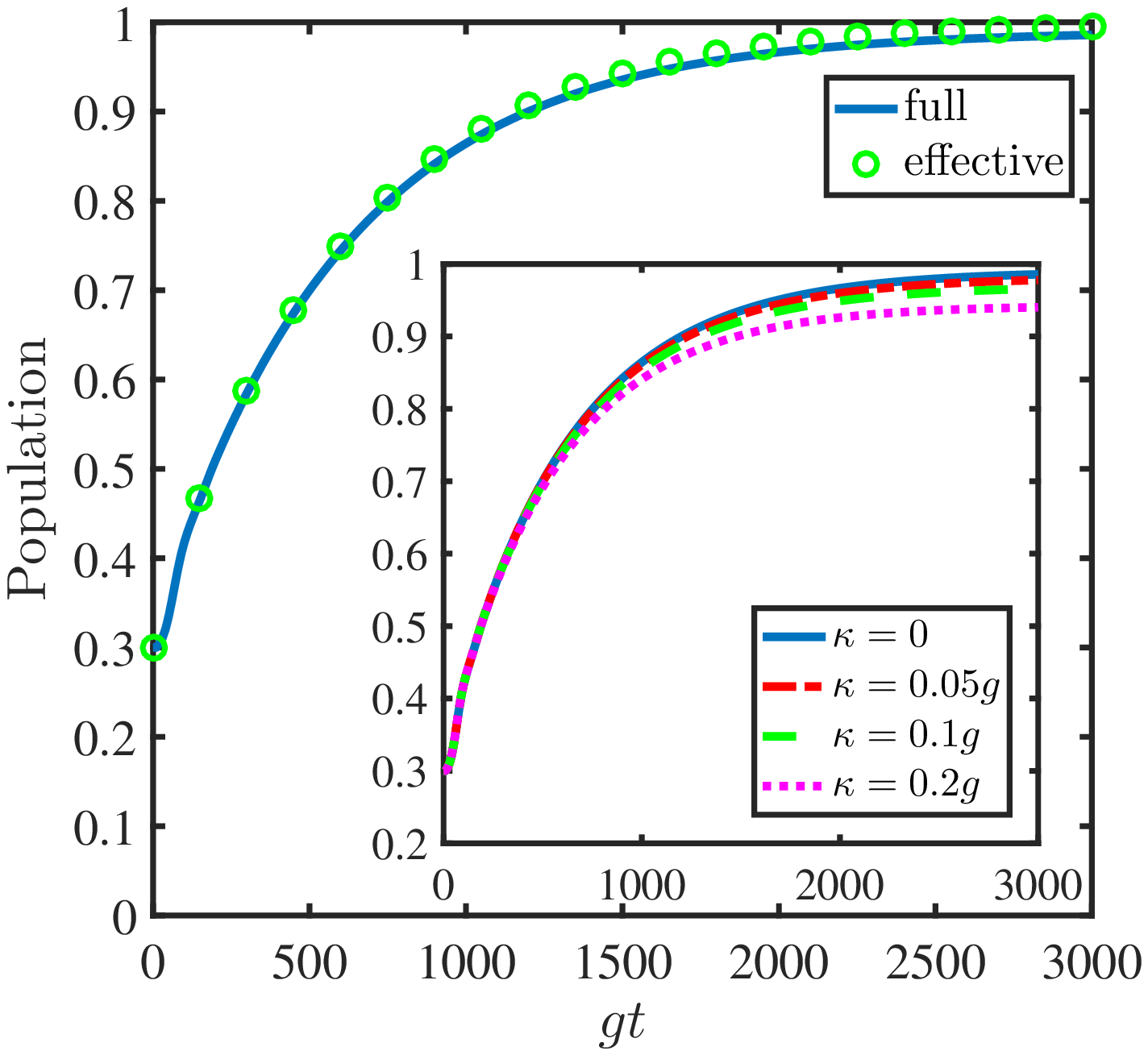}
 \centerline{(c)}
 \end{minipage}
\caption{(Color online) (a) Atomic level configuration and (b) effective transitions for generation of two-atom KLM state. (c) Populations of KLM state governed by full and effective master equation are plotted as functions of $gt$ and the inset shows the populations with different $\kappa$.   }
\label{effect2}
\end{figure}
Furthermore, our scheme not only can be used to prepare the antisymmetric entangled state $|S\rangle$, but also possesses the ability to produce some specific entanglement. For instance, the Knill-Laflamme-Milburn (KLM) state is a particular
form of entangled state, which is defined as
\cite{KMLnature}
$
|t_n\rangle=\frac{1}{\sqrt{1+n}}\sum_{j=0}^n|1\rangle^j|0\rangle^{n-j},
$
where the notation $|a\rangle^j$ means $|a\rangle|a\rangle...,j$ times.
 Employing the KLM states as ancillary resources,
one can improve the success probability of teleportation
gradually closed to unity with the increase of the
particle number in the KLM state. For the case of $n=2$, the two-particle KLM state reads
$|t_2\rangle=(|00\rangle_{12}+|10\rangle_{12}+|11\rangle_{12})/\sqrt{3}.$
To prepare the above state dissipatively, we remove the resonantly optical laser driving the transition between state $|0\rangle$ and state $|2\rangle$ of the second atom, set $\delta=\Omega_{MW}$, and keep a $\pi$-phase difference between the Rabi frequencies of the dispersive microwave fields acting on two atoms, as illustrated in Fig.~\ref{effect2}(a).
The effective Hamiltonian $H_{\rm{eff}}$ can be rewritten as
\begin{eqnarray}\label{effectH2}
H_{\rm{eff}}&=&\Omega_{MW}(|00\rangle-|11\rangle)(\langle10|-\langle01|)+\frac{\Omega}{\sqrt{2}}|D\rangle\langle 01|+{\rm{H.c.}}\nonumber\\
&&+\delta (|D\rangle\langle D|+|01\rangle\langle 01|+|10\rangle\langle 10|+2|11\rangle\langle11|).
\end{eqnarray}
The effective Lindbald operators are
$L_{{\rm eff}}^1=\sqrt{\gamma/2}|11\rangle\langle D|,$
$L_{{\rm eff}}^{2(3)}=\sqrt{\gamma/4}|10(01)\rangle\langle D|.$
The corresponding effective transitions of states are shown in Fig.~\ref{effect2}(b). The four ground states $|00\rangle,|10\rangle,|01\rangle,$ and $|11\rangle$ can transfer to each other owing to the effect of microwave field. Since the driving of the optical pumping laser with Rabi frequency $\Omega$, the transition $|01\rangle\leftrightarrow|D\rangle$ will always occur. Meanwhile, the spontaneous emission of the excited state $|D\rangle$ results in the states $|10\rangle,|01\rangle$ and $|11\rangle$ populated. Combining the detuning of microwave field and the condition $\delta=\Omega_{MW}$,
the target state $|t_2\rangle$ becomes the unique steady state of system. Ultimately, the system is stabilized into the two-atoms entangled KLM state.

In the Fig.~\ref{effect2}(c), the validity of the above analysis for the reduced system is proved by the coincidence of the populations of KLM state governed by full (solid line) and effective (empty circles) master equations. The parameters are same as the Fig.~\ref{model}(c) except for $\Omega=0.05g$. The inset plots the populations of KLM state governed by full master equation with different $\kappa$ and also reflects the robustness of the present scheme.

Finally, we discuss the experimental feasibility of the
scheme. Ref.~\cite{pra054302ref35} reported the projected limits for a Fabry-Perot cavity. The coupling coefficient $g/2\pi$ is $770~\rm{MHz}$. Based on the corresponding critical photon number and critical atom number, we obtain $(\kappa,\gamma)/2\pi=(21.7,2.6)~\rm{MHz}$. These parameters make the fidelity $F=\sqrt{\langle S|\rho(t)|S\rangle}$ and $F=\sqrt{\langle t_2|\rho(t)|t_2\rangle}$ reach $99.66\%$ and $99.75\%$, respectively, while other relevant parameters are selected as $\Omega=0.01g,~\Omega_{MW}=0.5\Omega$ and $\delta=\Omega_{MW}$. Moreover, in a microscopic optical resonator \cite{srref68}, the parameters of an atom interacting with an evanescent field are $(g,\kappa,\gamma)/2\pi=(70,5,1)~\rm{MHz}$,  which correspond to the fidelity $99.71\%$ and $99.77\%$ of our proposal. In another experiment \cite{pra054302ref32,pra054302ref34}, Cesium atoms are trapped inside a high finesse optical resonator, where the maximum single-photon Rabi frequency is $g/2\pi=34~\rm{MHz}$. The decay rate is $\gamma/2\pi=2.6~\rm{MHz}$, and the cavity field decays at a rate $\kappa/2\pi=4.1~\rm{MHz}$. Substituting these values into the steady-state solution of the full schemes of $|S\rangle$ and $|t_2\rangle$, the fidelity is acquired as $99.18\%$ and $99.19\%$, respectively.

In summary, we have systematically investigated the feasibility for preparing maximal entanglement of two $\Lambda$-type atoms via dissipation. The spontaneous emission of atoms is actively exploited, and the cavity decay is suppressed due to the quantum Zeno dynamics. The dispersive microwave field efficiently eliminates the energy degeneracy of ground state, resulting in a unique steady solution of system. An effective master equation for describing the evolution of system is established, which is then confirmed by strictly numerical simulations. Therefore, our scheme guarantees that a high-fidelity steady
entangled state can be prepared for a wide range of decoherence parameters. We believe that our work may provide a new opportunity
for entanglement preparation experimentally.

This work is supported by National Natural Science Foundation of China (NSFC) under Grants No. 11534002,
No. 61475033, No. 11774047, and Fundamental Research Funds for the Central Universities under Grants No. 2412016KJ004.

\bibliographystyle{apsrev4-1}
\bibliography{MMF}

\end{document}